# The *q*-Diode

R. V. Ramos

rubens.ramos@ufc.br

*Lab. of Quantum Information Technology, Department of Teleinformatic Engineering – Federal University of Ceara - DETI/UFC, C.P. 6007 – Campus do Pici - 60455-970 Fortaleza-Ce, Brazil.*

*Abstract*

The present work introduces the new function $R_{q,Q}(z)$, solution of the equation $R_{q,Q}(z) \times_Q \exp_q(R_{q,Q}(z)) = z$. It is shown this new function can be used to construct a new disentropy as well it is used to model the *q*-diode, a hypothetical electronic device whose electrical current depends *q*-exponentially on the voltage between its terminals.

## 1. Introduction

The Lambert *W* function is an important elementary mathematical function that finds applications in different areas of mathematics, computer Science and physics [1-6]. The Lambert *W* function is defined as the solution of the equation

$$W(z)e^{W(z)} = z. \qquad (1)$$

In the interval $-1/e \leq x \leq 0$ there exist two real values of $W(z)$. The branch for which $W(x) \geq -1$ is the principal branch named $W_0(z)$ while the branch satisfying $W(z) \leq -1$ is named $W_{-1}(z)$. For $x \geq 0$ only $W_0(z)$ is real and for $x < -1/e$ there are not real solutions. The point $(z_b = -1/e, W(z_b) = -1)$ is the branch point where the solutions $W_0$ and $W_{-1}$ have the same value.

On the other hand, the *q*-exponential function proposed by Tsallis [7] is given by

$$e_q^z = \begin{cases} e^z & q=1 \\ \left[1+(1-q)z\right]^{1/(1-q)} & q \neq 1 \ \& \ 1+(1-q)z \geq 0 \\ 0 & q \neq 1 \ \& \ 1+(1-q)z < 0 \end{cases} \qquad (2)$$

Using Tsallis *q*-exponential (2) in the Lambert equation (1), one has the Lambert-Tsallis equation

$$W_q(z) e_q^{W_q(z)} = z \tag{3}$$

whose solutions are the Lambert-Tsallis $W_q$ functions [8]. Using the definition of $exp_q$ given in eq. (2) in eq. (3), the $W_q$ function can be found solving the equation [8]

$$x(r+x)_+^r = r^r z, \tag{4}$$

where $x = W_{\frac{(r-1)}{r}}(z)$, $r = 1/(1-q)$ and $(x)_+ = \max\{x,0\}$. When $q = 1$, one has $e_1(z) = e^z$ and, consequently, $W_1(z) = W(z)$. For example, for $q = \{2, 3, 3/2, 1/2\}$ one has the following Lambert-Tsallis $W_q$ upper branches

$$W_2(z) = \frac{z}{z+1}, \qquad z > -1, \tag{5}$$

$$W_3(z) = z\sqrt{z^2+1} - z^2 \quad (z \geq 0). \tag{6}$$

$$W_{3/2}^+(z) = \frac{2(z+1) + 2\sqrt{2z+1}}{z}, \qquad z > -1/2, \tag{7}$$

$$W_{1/2}^+(z) = \frac{\left[3\sqrt[3]{2z + \sqrt{\left(2z+\frac{8}{27}\right)^2 - \frac{64}{729}} + \frac{8}{27}} - 2\right]^2}{9\sqrt[3]{2z + \sqrt{\left(2z+\frac{8}{27}\right)^2 - \frac{64}{729}} + \frac{8}{27}}}, \qquad z \geq -0.29629, \tag{8}$$

Figure 1 shows the plot of $W_{q=3/2}$ versus *z*.

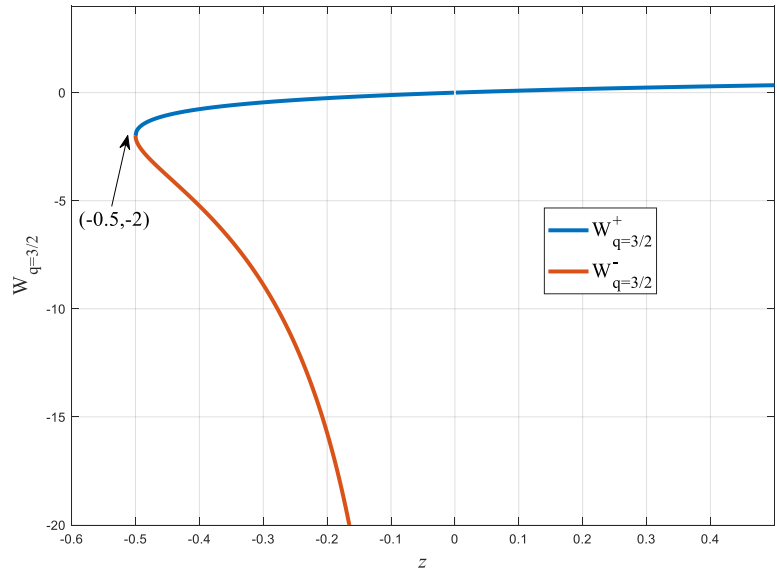

Fig.1. $W_{q=3/2}$ versus $z$.

More details about the Lambert-Tsallis function and its applications can be found in [8-15].

In order to handle with the $exp_q$ function, one has to use the $q$-operations. The important ones used in this work are:

$$a \times_q b = \max\left\{\left[a^{(1-q)} + b^{(1-q)} - 1\right]^{1/(1-q)}, 0\right\} \equiv \left[a^{(1-q)} + b^{(1-q)} - 1\right]_+^{1/(1-q)} \tag{9}$$

$$\left(e_q^x\right)^\alpha = e_{1-(1-q)/\alpha}^{\alpha x}. \tag{10}$$

## 2. The $R_{q,Q}$ function

In this section a new function is introduced. It is named $R_{q,Q}$ function and it is the solution of the following equation

$$R_{q,Q}(z) \times_Q e_q^{R_{q,Q}(z)} = z. \tag{11}$$

Equation (11) is the Lambert-Tsallis equation using the $q$-product operation. Obviously, $R_{q,Q=1}(z) = W_q(z)$. Using (2) and (9) in (11) one gets

$$R_{q,Q}^{1-Q}(z) + \left[1+(1-q)R_{q,Q}(z)\right]^{\frac{1-Q}{1-q}} - \left(z^{1-Q}+1\right) = 0. \tag{12}$$

The general solutions of (12) will be published elsewhere. Here, the important case for introduction of a new disentropy and the $q$-diode modelling is $Q = q$. In this case eq. (12) is reduced to

$$R_{q,q}^{1-q}(z) + (1-q)R_{q,q}(z) - z^{1-q} = 0. \tag{13}$$

For example, for $q = 2$ and $q = 1/2$ one has

$$R_{2,2}(z) = -\frac{1}{2z} \pm \frac{1}{2}\sqrt{\frac{1}{z^2}+4}, \tag{14}$$

$$R_{1/2,1/2}(z) = 2\left(z^{1/2}+1\right) - 2\sqrt{2z^{1/2}+1}. \tag{15}$$

Figure 2 shows the plot of the parts of the functions $R_{2,2}$ and $R_{1/2,1/2}$ that obey eq. (11).

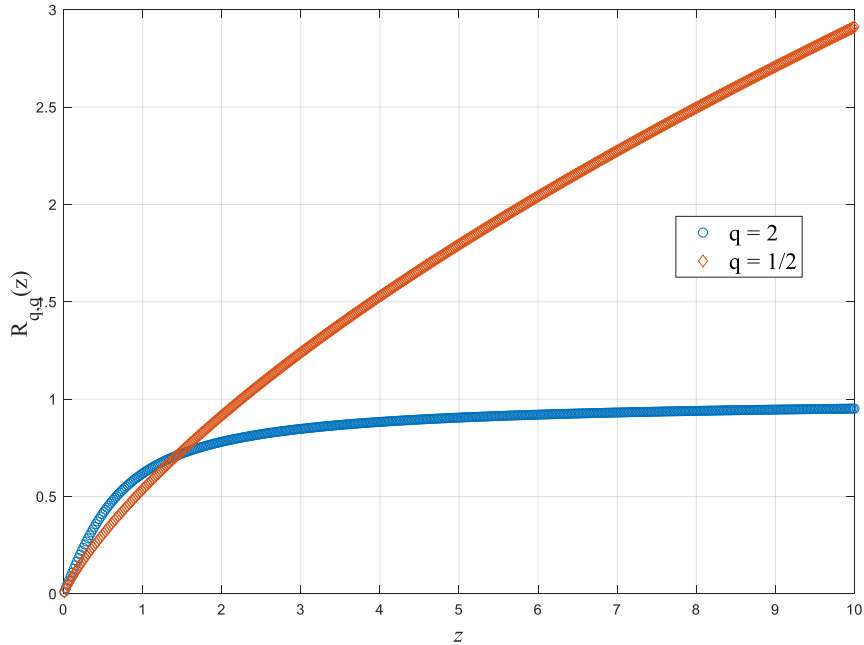

Fig. 2. $R_{q,q}(z)$ versus $z$ for $q = ½$ and $q = 2$.

## 3. Disentropy

The disentropy based on the Lambert and Lambert-Tsallis functions and its applications in quantum and classical information theory, image processing and black hole, among others, have been discussed in [8-14]. Taking the $\log_q$ in both sides of eq. (11) with $q = Q$, one gets

$$\log_q(z) = R_{q,q}(z) + \log_q\left[R_{q,q}(z)\right]. \tag{16}$$

Hence, Tsallis $q$-entropy can be written as

$$S_q = \sum_i p_i^q \log_q(p_i) = \sum_i p_i^q R_{q,q}(p_i) + \sum_i p_i^q \log_q\left[R_{q,q}(p_i)\right]. \tag{17}$$

The term

$$D_q = \sum_i p_i^q R_{q,q}(p_i) \tag{18}$$

is a disentropy. It can be shown it is maximal for delta distribution and minimal for a uniform distribution. Its quantum version is

$$D_q(\rho) = \sum_i \lambda_i^q R_{q,q}(\lambda_i) \tag{19}$$

where $\lambda_i$ is the $i$-th eigenvalue of the density matrix $\rho$. The disentropy based on the $R_{q,q}$ function can be used in the same problems that the disentropy based on the Lambert-Tsallis function is used. For example, it can be used to measure the disentanglement of bipartite of qubit states [8]. Figure 3 shows the behaviour of $D_q$ for the distribution $\{p, 1-p\}$ using the values $q = 0.5$, $q = 1$ and $q = 2$.

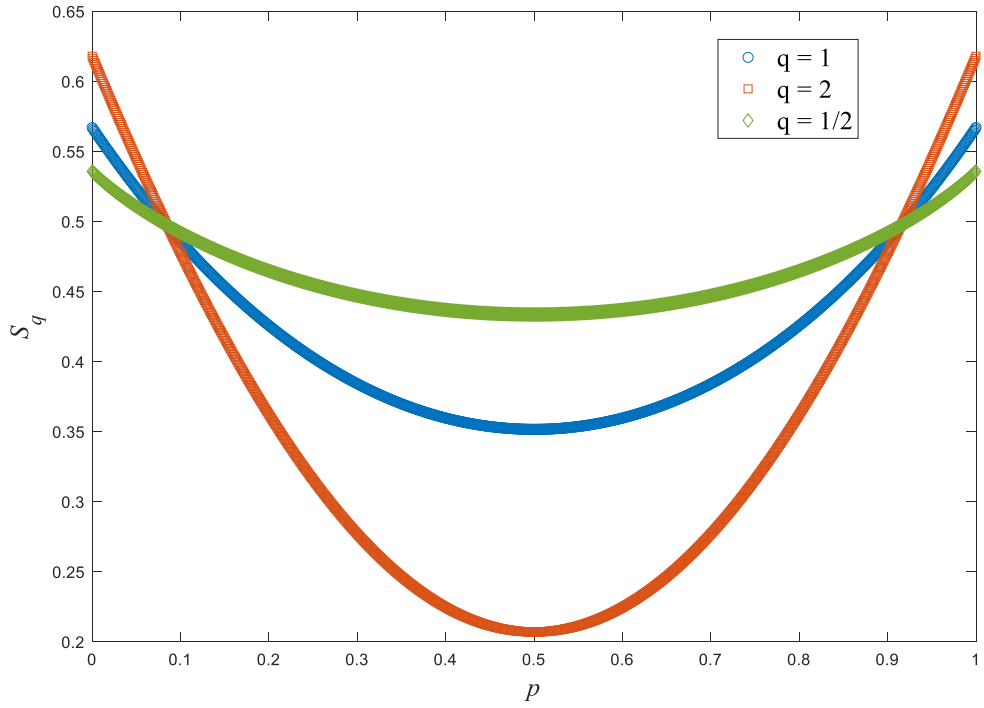

Fig. 3. Disentropy of the distribution $\{p, 1-p\}$ versus $p$ for $q \in [0.5, 1, 2]$.

## 4. The $q$-Diode

For a semiconductor diode that obeys the Schockley's model, the relation between current and voltage is given by

$$I = I_s e^{\frac{V_D}{\eta V_T}}, \qquad (20)$$

where $I_s$ is the saturation current of the diode, $V_D$ is the voltage between the diode terminals, $V_T = kT/q_e$ ($q_e$ – electron charge, $k$ – Boltzman constant, $T$ - temperature in Kelvin) and, finally, $\eta$ is the diode ideality factor ($1 < \eta < 2$ for silicon diodes). Figure 4 shows the very basic electrical circuit composed by a power supply, a resistor and the diode.

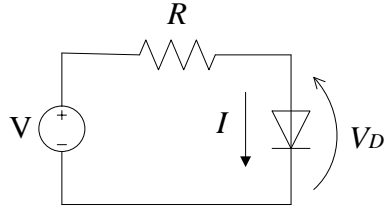

Fig. 4 – Electrical circuit with a resistor and a diode.

The current that flows through the diode in the circuit shown in Fig. 3 is given by

$$I = I_s e^{\frac{V-RI}{\eta V_T}}. \tag{21}$$

Using the Lambert $W$ function in (21) one gets the following relation between electrical current ($I$) and power supply voltage ($V$)

$$I = \frac{\eta V_T}{R} W\left(\frac{I_s R}{\eta V_T} e^{\frac{V}{\eta V_T}}\right). \tag{22}$$

The $q$-diode, by its turn, is defined as the hypothetical device whose relation between current and voltage between its terminals ($V_D$) is given by

$$I = I_s e_q^{\frac{V_D}{\eta V_T}}. \tag{23}$$

Using the $q$-diode in the circuit shown in Fig. 4, the value of the electric current flowing through the diode is given by

$$I = I_s e_q^{\frac{V-IR}{\eta V_T}}. \tag{24}$$

Using the *q*-operations in (20) one gets

$$I = I_s e_q^{\frac{V-IR}{\eta V_T}} \Rightarrow \frac{IR/\eta V_T}{I_s R/\eta V_T} = e_q^{\frac{V}{\eta V_T}} \div_q e_q^{\frac{IR}{\eta V_T}} \Rightarrow \frac{IR/\eta V_T}{I_s R/\eta V_T} \times_q e_q^{\frac{IR}{\eta V_T}} = e_q^{\frac{V}{\eta V_T}}. \tag{25}$$

Now, using the function $R_{q,q}$ in (25), after some algebra one gets the following solutions for the electrical current *I*, for $q = 2$ and $q = 0.5$,

$$I_2(V) = \frac{\eta V_T}{R}\left[ -\frac{1}{2e_q^{\frac{V}{\eta V_T}}} + \frac{1}{2}\sqrt{\frac{1}{\left(e_q^{\frac{V}{\eta V_T}}\right)^2} + \frac{4\eta V_T}{I_s R}} \right], \tag{26}$$

$$I_{1/2}(z) = \frac{\eta V_T}{R}\left[ 2\left(\left(e_q^{\frac{V}{\eta V_T}}\right)^{1/2} + \frac{\eta V_T}{I_s R}\right) - 2\sqrt{2\frac{\eta V_T}{I_s R}\left(e_q^{\frac{V}{\eta V_T}}\right)^{1/2} + \left(\frac{\eta V_T}{I_s R}\right)^2} \right]. \tag{27}$$

One may note that (26) and (27) are, respectively, equal to (14) and (15) when ($\eta V_T/I_s R$) = 1. In Fig. 5 one can see the comparison between the cases $q = 1$, $q = 0.75$ and $q = ½$. The smaller the value of *q* the slower is the growth of the current. The *q*-diode with $q > 1$ operates at very low voltage since $\exp_q(x)$ goes too fast to zero. For example, for $q = 1.25$, one must have $V/(\eta V_T) < 4$ ($V < \sim 0.1$mV).

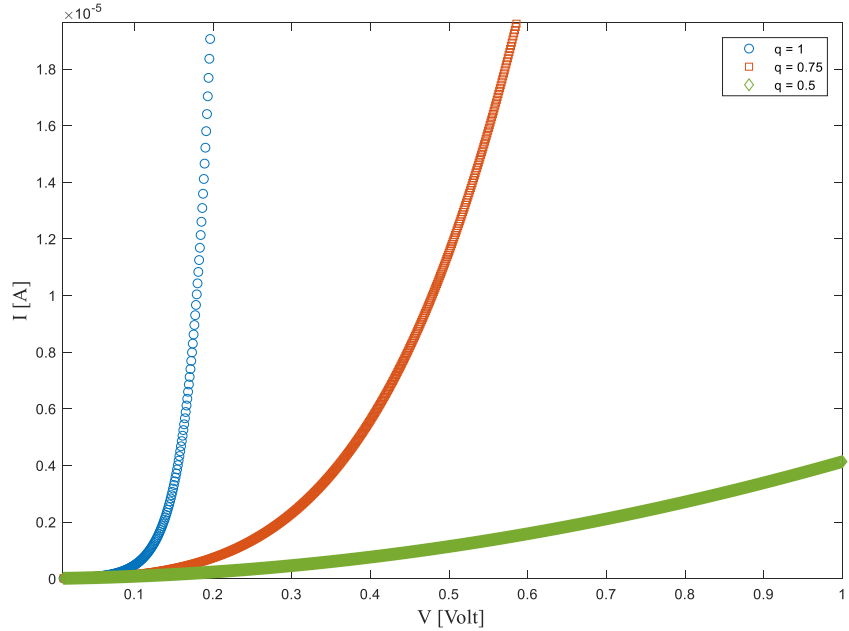

Fig. 5 – $q$-Diode current versus voltage curve for $q \in [0.5, 0.75, 1]$.

## 5. Conclusions

Initially, the present work introduced the solutions of the equation $R_{q,Q}(z) \times_Q \exp_q(R_{q,Q}(z)) = z$ and showed two applications of the function $R_{q,Q}(z)$: 1) It was used to construct a new disentropy formula. This new disentropy can be applied in a large variety of problems in physics and engineering. A comparison between the disentropy based on the $R_{q,q}$ function and the disentropy based on the Lambert-Tsallis $W_q$ function is a question for future investigation. 2) It was used to model the $q$-diode. Basically, compared to the classical diode, the $q$-diode with $q > 1$ has to operate with lower voltage while the $q$-diode with $q < 1$ requires a larger voltage. Since, the $q$-diode shows the nonlinear behaviour (between $I$ and $V$) it can be used in an electronic circuit as modulator or mixer, for example. Which values of $q$ will result in a $q$-diode that can be realized physically is still a problem to be investigated.

## Acknowledgements

This study was financed in part by the Coordenação de Aperfeiçoamento de Pessoal de Nível Superior - Brasil (CAPES) - Finance Code 001, and CNPq via Grant no. 307184/2018-8. Also, this work was performed as part of the Brazilian National Institute of Science and Technology for Quantum Information.